\documentclass[preprint,showpacs,preprintnumbers,amsmath,amssymb]{revtex4}
\usepackage{amsfonts, amssymb,amsmath}
\usepackage{graphicx}% Include figure files
\usepackage{dcolumn}% Align table columns on decimal point
\usepackage{bm}% bold math

\begin{document}

%------------------------------------------------------------------------------------
\title{Superluminal Pulse Propagation in a One-sided Nanomechanical Cavity System}
\author{Devrim Tarhan}
\affiliation{ Department of Physics, Harran University, 63300
Osmanbey Yerle\c{s}kesi, \c{S}anl\i{}urfa, Turkey}
\email{dtarhan@harran.edu.tr}

\date{\today}
%-------------------------------------------------------------------------------
\begin{abstract}
We investigate the propagation of a pulse field in an
optomechanical system. We examine the question of advance of the
pulse under the conditions of electromagnetically induced
transparency in the mechanical system contained in a high quality
cavity. We show that the group delay can be controlled by the
power of the coupling field. The time delay is negative which
corresponds to superluminal light when there is a strong coupling
between the nano-oscillator and the cavity.

\end{abstract}
 \maketitle
%%%%%%%%%%%%%%%%%%%%%%%%%%%%%%%%%%%%%%%%%%%%%%%%%%%%%%%%%%%%%%%%%%%%%%%%%%%%%%%%%%%%%%%%%%%%%%%
%42.50.Gy   Effects of atomic coherence on propagation, absorption, and amplification of light;
%           electromagnetically induced transparency and absorption
%
%42.50.Ct Quantum description of interaction of light and matter; related experiments
%
%42.50.Wk Mechanical effects of light on material media, microstructures and particles (see also 87.80.Cc Optical trapping in biology and medicine)
%
%
%%%%%%%%%%%%%%%%%%%%%%%%%%%%%%%%%%%%%%%%%%%%%%%%%%%%%%%%%%%%%%%%%%%%%%%%%%%%%%%%%%%%%%%%%%%%%%%%
%-------------------------------------------------------------------------------
%%%%%%%%%%%%%%%%%%%%%%%%%%%%%%%%%%%%%%%%%%%%%%%%%%%%%%%%%%%%%%%%%%%%%%%%%%%%%%%%%%%%%%%%%%%%%%%%%%%%%%%%%%%%%%%%%%%%%%%%%%%%%%%%%%%%%%%%%%%%%%%%%%%%%%%%%%%%%%%%%%%%%%%%%%%%%
%%%%%%%%%%%%%%%%%%%%%%%%%%%%%%%%%%%%%%%%%%%%%%%%%%%%%%%%%%%%%%%%%%%%%%%%%%%%%%%%%%%%%%%%%%%%%%%%%
\section{Introduction}

The experimental accomplishment of ultraslow light
\cite{slowlight-exp1} in ultracold atoms by using
electromagnetically induced transparency\cite{eit1} (EIT),
inspired appealing applications
\cite{slowlight-apps1,slowlight-apps2,slowlight-apps3,slowlight-apps4}.
Besides the ultraslow light, fast light effects ($ v_g > c \,\, or
\, \, v_g \, \,$ is negative) were observed experimentally by
using a gain-assisted linear anomalous dispersion medium
\cite{kuzmich,kuzmich1} and coherent population oscillations
\cite{bigelow}. Slow light has been also observed in
nanomechanical systems very recently \cite{painter}. In this
experiment, the transmitted signal delay can be tuned with control
beam power in a side-coupled cavity system. There has been great
interest in nanomechanical optical systems more recently
\cite{painter1,favero,schliesser,aspel,meystre,kippenberg,nunnenkamp,schwab,agarwal1}.
At the same time light storage in an optical waveguide coupled to
an optomechanical crystal array has been suggested that
information is coherently transferred to mechanical vibrations of
the array \cite{painter1}. On the other hand, quite recently
optomechanical analogy of EIT in a high-quality cavity with
nanomechanical mirror has been considered independently
\cite{agarwal1}. The experimental analogue of electromagnetically
induced transparency has been demonstrated in a room temperature
cavity optomechanics setup formed by a thin semitransparent
membrane within a Fabry-Perot cavity \cite{vitali1} and detailed
experimental studies on optomechanical light storage in a silica
microresonator have been done very recently \cite{fiore,fiore1}.
And also EIT in a cavity optomechanical system with an atomic
medium has been studied \cite{zhou}. Furthermore, tunable group
delay of the pulse in a cavity optomechanical system with a
Bose-Einstein condensate \cite{kadizhu} and in a nanomechanical
resonator-superconducting microwave cavity systems \cite{kadizhu1}
have been studied theoretically very recently. Moreover,
nonclassical states in optomechanical systems can be generated
experimentally. Entangled states can be prepared between the probe
light field and the oscillating nano-mirror (see
Ref.\cite{marquardt}  and references therein). It has shown
experimentally that entanglement between an optical cavity field
mode and a macroscopic vibrating mirror can be generated by means
of radiation pressure \cite{vitali}. A theoretical work has shown
that squeezing of the movable mirror can be achieved by the
injection of squeezed vacuum light and a laser recently
\cite{sumei}.

In this work,  we investigate the time delay of the weak probe
field at the probe resonance in a high quality cavity
nanomechanical mirror under the action of coupling laser. We
investigate the propagation of a pulse field in an optomechanical
system and examine the question of advance of the pulse under the
conditions of electromagnetically induced transparency in the
mechanical system contained in a high quality cavity. We show that
the group delay can be controlled by the power of the coupling
field. Before this work, there were some theoretical studies on
the tunable group delay of the pulse in a cavity optomechanical
system with a Bose-Einstein condensate \cite{kadizhu} and in a
nanomechanical resonator-superconducting microwave cavity systems
\cite{kadizhu1}. However our group delay results are in
disagreement with a theoretical study \cite{kadizhu1} for the
nanomechanical resonator-superconducting microwave cavity systems.

%%%%%%%%%%%%%%%%%%%%%%%%%%%%%%%%%%%%%%%%%%%%%%%%%%%%%%%%%%%%%%%%%%%%%%%%%%%%%%%%%%%%%%%%%%%%%%%%%%%%%%%%%%%%%%%%%%%%%%%%%%%%%%%%%%%%
%%%%%%%%%%%%%%%%%%%%%%%%%%%%%%%%%%%%%%%%%%%%%%%%%%%%%%%%%%%%%%%%%%%%%%%%%%%%%%%%%%%%%%%%%%%%%%%%%%%%%%%%%%%%%%%%%%%%%%%%%%%%%%%%%%%%
\section{Model System}
\label{sec:model}

Let us consider the same model in Ref. (\cite{agarwal1}). The
pump-probe response of a nanomechanical oscillator of frequency
$\omega_m$ is coupled to a high quality cavity via radiation
pressure effects \cite{agarwal1}. As shown in Fig.(1) the system
is driven by a coupling field of frequency $\omega_c$ and the
probe field has frequency $\omega_p$.
\begin{figure}[htbp]
\centering{\vspace{0.5cm}}
\includegraphics[width=8.5cm]{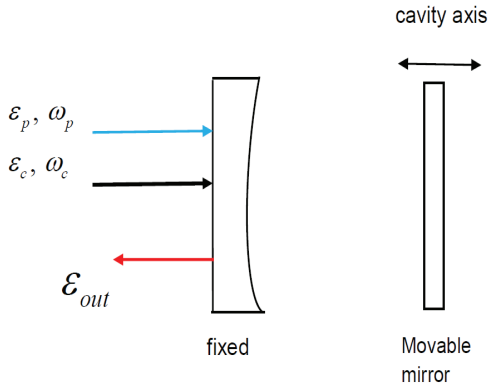}
\caption{(Color online) Schematic of a optomechanical system
coupled to a high quality cavity via radiation pressure effects
adapted from Ref.{\cite{agarwal1}}.} \label{fig1}
\end{figure}
The Hamiltonian of this system is \cite{agarwal1}:
\begin{eqnarray} \label{Ham}
H  &=& \hbar \omega_0  c^{\dagger} c + (\frac{p^2}{2m}+\frac{1}{2}m \omega_m^2 q^2 )+ i\hbar (c^{\dagger}  \varepsilon_c e^{-i\omega_c t}-c  \varepsilon_c^\star e^{i\omega_c t}) \nonumber \\
&-& i\hbar (c^{\dagger}\varepsilon_p e^{-i\omega_p t}-c
\varepsilon_p^\star e^{i\omega_p t})-\chi_0 c^{\dagger} c q,
\end{eqnarray}
where $\hat{c}^{\dagger}$ and $\hat{c}$ are the creation and
annihilation operators in the cavity respectively; $p$ and $q$ are
the momentum and position operators of the nanomechanical
oscillator respectively; the amplitudes of the pump and probe
field are introduced by $|\varepsilon_c|=\sqrt{2\kappa P_c/\hbar
\omega_c}$ and $|\varepsilon_p|=\sqrt{2\kappa P_p/\hbar \omega_p}$
respectively. $P_c$ and $P_p$ are the pump and probe powers
respectively. $\chi_0=\hbar \omega_0/L$ is the coupling constant
between the cavity field and the movable mirror, where $L$ is the
cavity length. Quantum fluctuations are not taken into account
because we are dealing with the main response of the system. We
use the mean field assumption $ \langle Q c \rangle = \langle Q
\rangle \langle c \rangle $ in order to derive
Eq.(\ref{Heisenberg}) \cite{agarwal2}. The mean value equations
are found by Heisenberg equation of motion($\langle \dot{O}
\rangle= \langle 1/i \hbar [O,H] \rangle $)by adding the damping
and the noise terms \cite{agarwal1}:

\begin{eqnarray} \label{Heisenberg}
\langle \dot{q} \rangle  &=& \frac{\langle p \rangle}{m}  \nonumber \\
\langle \dot{p} \rangle &=& -m \omega_m^2 \langle q \rangle +
\chi_0
 \langle \hat{c}^{\dagger} \rangle \langle \hat{c} \rangle - \gamma_m \langle p \rangle \nonumber \\
 \langle \dot{\tilde{c}} \rangle=&-&[\kappa +
i(\omega_0-\omega_c-\frac{\chi_0}{\hbar} \langle q \rangle)]
\langle \tilde{c} \rangle .
\\&+&\varepsilon_c+\varepsilon_p e^{-i(\omega_p-\omega_c)t}
\nonumber
\end{eqnarray}
At the rotating frame at the frequency $\omega_c$ transformed
operator $ \langle \tilde{c}(t) \rangle= \langle c(t) \rangle e^{i
\omega_c t}$. If we use input-output relations \cite{collet}, we
obtain
\begin{equation}
\frac{1}{\sqrt{2 \kappa}} \{ \varepsilon_{out}(t) +
\varepsilon_{p}e^{-i \omega_p t} + \varepsilon_{c}e^{-i \omega_c
t} \}  = \sqrt{2 \kappa} \langle c\rangle.
\end{equation}
In this system for the cavity decay rate $2\kappa$ is used. We use
the following anzats\cite{boyd} in terms of probe field so as to
solve Eq.(\ref{Heisenberg}):
\begin{eqnarray}
q(t)  &=& q_0 + q_+ \varepsilon_p e^{-i \delta t} + q_- \varepsilon_p^* e^{i \delta t}  \nonumber \\
p(t)  &=& p_0 + p_+ \varepsilon_p e^{-i \delta t} + p_- \varepsilon_p^* e^{i \delta t}.  \\
\tilde{c}(t)  &=& \tilde{c}_0 + \tilde{c}_+ \varepsilon_p e^{-i
\delta t} + \tilde{c}_- \varepsilon_p^* e^{i \delta t} \nonumber
\label{anzat}
\end{eqnarray}
The field $\varepsilon_p$ with frequency $\omega_p$ is much weaker
than the pump field $\varepsilon_c$. We obtain the steady state
and first order solutions. The zeroth order and first order
equations according to $\varepsilon_p$ are:
\begin{eqnarray} \label{sol1}
p_0 = 0 \nonumber \\
-i(\omega_p-\omega_c)q_+ = \frac{p_+}{m} \\
i(\omega_p-\omega_c)q_- = \frac{p_-}{m}  \nonumber
\end{eqnarray}
\begin{eqnarray} \label{sol2}
-m \omega_m^2 q_0 + \chi_0 |\tilde{c}_0|^{2} - \gamma_m p_0 = 0 \nonumber \\
-i(\omega_p-\omega_c)p_+  = -m \omega_m^2 q_+ + \chi_0(\tilde{c}_0^\star \tilde{c}_+ + \tilde{c}_0 \tilde{c}_-^\star)  \\
i(\omega_p-\omega_c)p_-  = -m \omega_m^2 q_- +
\chi_0(\tilde{c}_0^\star \tilde{c}_- + \tilde{c}_+^\star
\tilde{c}_0 ) \nonumber. \label{anzat}
\end{eqnarray}
\begin{eqnarray} \label{sol3}
-[\kappa + i(\omega_0 - \omega_c)]\tilde{c}_0 + i \frac{\chi_0}{\hbar} q_0 \tilde{c}_0 + \varepsilon_c = 0 \nonumber \\
-i(\omega_p-\omega_c)\tilde{c}_+  = -[\kappa + i(\omega_0 - \omega_c)]\tilde{c}_+ + i \frac{\chi_0}{\hbar}(q_0 \tilde{c}_+ + q_+ \tilde{c}_0)+1 \\
i(\omega_p-\omega_c)\tilde{c}_-  = -[\kappa + i(\omega_0 -
\omega_c)]\tilde{c}_-  + i \frac{\chi_0}{\hbar}(q_0 \tilde{c}_- +
q_- \tilde{c}_0) \nonumber. \label{anzat}
\end{eqnarray}
When we use steady state equations in Eq.(\ref{sol1}),
Eq.(\ref{sol2}), and Eq.(\ref{sol3}), we get the steady state
solutions $ \langle \tilde{c}_0 \rangle = \varepsilon_c/(\kappa+i
\Delta)$, $ \langle q_0 \rangle = \chi_0 |\varepsilon_c|^2/[m
\omega_m^2 (\kappa^2+\Delta^2)]$, and $\Delta=\omega_0 - \omega_c
- \chi_0 \,\, q_0 / \hbar$. If we use first order equations in
Eq.(\ref{sol1}), Eq.(\ref{sol2}), and Eq.(\ref{sol3}), we get the
$\tilde{c}_+$
\begin{eqnarray} \label{cplus}
\tilde{c}_+ = \frac{ m(\delta^2 - \omega_m^2 + i \gamma_m
\delta)[\kappa - i(\Delta + \delta)]-i \alpha }{m(\delta^2 -
\omega_m^2 + i \gamma_m \delta)[\kappa + i(\Delta-\delta)][\kappa
- i(\Delta+\delta)] +2 \Delta \alpha }
\end{eqnarray}
where $\delta=\omega_p-\omega_c$, and $\alpha=\chi_0^2
|\tilde{c}_0|^2/\hbar$ .
%%%%%%%%%%%%%%%%%%%%%%%%%%%%%%%%%%%%%%%%%%%%%%%%%%%%%%%%%%%%%%%%%%%%%%%%%%%%%%%%%%%%%%%%%%%%%%%%%%%%%%%%%%%%%%%%%%%%%%%%%%%%%%%%%%%%
%%%%%%%%%%%%%%%%%%%%%%%%%%%%%%%%%%%%%%%%%%%%%%%%%%%%%%%%%%%%%%%%%%%%%%%%%%%%%%%%%%%%%%%%%%%%%%%%%%%%%%%%%%%%%%%%%%%%%%%%%%%%%%%%%%%%
\section{Results and Discussions}
\label{sec:results}

We can write the output field $\varepsilon_{out}(t)  =
\varepsilon_{out0} + \varepsilon_{out+} \varepsilon_p e^{-i \delta
t} + \varepsilon_{out-} \varepsilon_p e^{i \delta t}$ \cite{boyd}.
Inserting this to the input-output relation and compare the first
order according to the $\varepsilon_p$, we will get $(
\varepsilon_{out+} + 1)=2 \kappa \tilde{c}_+$. $2 \kappa
\tilde{c}_+$ is component of the output field. The output field is
given by
\begin{eqnarray} \label{eout1}
\varepsilon_{out+} = 2 \kappa \tilde{c}_+ - 1.
\end{eqnarray}
On the other hand, the output field can be written as
$\varepsilon_{out}(t)  = \varepsilon_{out+} \varepsilon_p e^{-i
\delta t}$. The amplitude of the output field is
\begin{equation} \label{eout1}
\varepsilon_{out+} = |T| \, e^{i \phi( \omega_p)}.
\end{equation}
The transmission is $T=1$ at resonance at
$\omega_p=\omega_c+\omega_m \equiv \overline{\omega}$. If we
expand $\phi(\omega_p)$ around $\overline{\omega}$, we will get
for the phase $\Phi$
\begin{equation} \label{phase1}
\Phi(\omega_p) = \Phi(\overline{\omega}) + (\omega_p -
\overline{\omega}) \frac{\partial \Phi}{\partial \omega_p} \mid
_{\overline{\omega}}.
\end{equation}
The transmitted probe pulse can be expressed as $|T|
\,\varepsilon_p e^{-i \omega_p t} e^{i
\Phi(\overline{\omega})}e^{-i (\omega_p -
\overline{\omega})\frac{\partial \Phi}{\partial \omega_p} \mid
_{\overline{\omega}}}$, where $\Phi(\overline{\omega})=0$ at
resonance. Combing with the $e^{-i \omega_p(t - \tau)}$, the
transmitted probe pulse peaks at $t=\tau$, where $\tau$ is the
pulse delay. Then the time delay of the pulse can defined as the
following:
\begin{equation}
\tau =  [ \frac{\partial \Phi}{\partial \omega_p} ] \mid
_{\overline{\omega}}.
\end{equation}
Phase of the output field can be found as
\begin{equation} \label{phase2}
\Phi = \frac{1}{2i} \ln(
\frac{\varepsilon_{out+}}{\varepsilon_{out+}^\star}).
\end{equation}
The time delay of the pulse can easily calculated from
Eq.(\ref{phase2}):
\begin{equation}
\tau = \texttt{Im} [\frac{1}{\varepsilon_{out+}}\frac{\partial
 \varepsilon_{out+}}{\partial \omega_p}] |_{\overline{\omega}} .
\end{equation}
In the absence of the coupling field the time delay is
$\tau=2/\kappa$. For numerical work we use experimental parameters
\cite{aspel}: the length of the cavity $L=25$ mm, the wavelength
of the laser $\lambda=2\pi c/\omega_c=1064$ nm, $m=145$ ng,
$\kappa=2\pi \times 215$ kHz, $\omega_m=2\pi \times 947$ kHz,
$\gamma=2\pi\times 141$ Hz, and mechanical quality factor
$Q=\omega_m/\gamma_m=6700$. If $\delta=\pm \omega_m$ and
$\Delta=\omega_m$ or in the sideband resolved limit $\omega_m \gg
\kappa$, the coupling between the nano-oscillator and the cavity
is the strongest. The real and the imaginer parts of the ($2
\kappa \tilde{c}_+$) represent the absorptive and dispersive
behavior, respectively.
\begin{figure}[htbp]
\centering{\vspace{0.5cm}}
\includegraphics[width=3.25in]{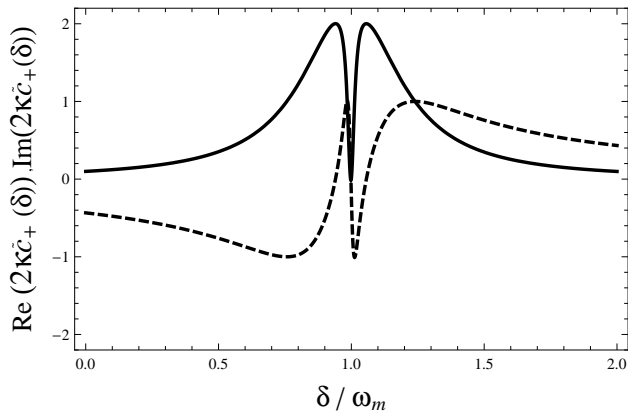}
\caption{ The real (solid) and the imaginary (dashed) parts of the
($2 \kappa \tilde{c}_+$) as a function of $\delta / \omega_m$ for
input coupling laser power of $P_c = 1$ mW. The parameters used
are the length of the cavity $L=25$ mm, the wavelength of the
laser $\lambda=2\pi c/\omega_c=1064$ nm, $m=145$ ng, $\kappa=2\pi
\times 215$ kHz, $\omega_m=2\pi \times 947$ kHz,
$\gamma=2\pi\times 141$ Hz, and mechanical quality factor
$Q=\omega_m/\gamma_m=6700$, and $\Delta=\omega_m$. This figure is
the same as in Ref.(\cite{agarwal1})} \label{fig2}
\end{figure}
%----------------------------------------------------

As seen in Fig.(\ref{fig2}) the imaginary part of the ($2 \kappa
\tilde{c}_+$) exhibits dispersive behavior however the slope of
the curve is negative. Therefore the group delay will become
negative. In Fig.(\ref{fig2}) we show the real and imaginary part
of the ($2 \kappa \tilde{c}_+$). Under the conditions of
electromagnetically induced transparency in the mechanical system
contained in a high quality cavity the system gives rise to
anomalous dispersion for the probe field which is in accord with
Ref.(\cite{agarwal1}). Because the slop of the curve in
Fig.(\ref{fig2}) is negative as well as in Ref.(\cite{agarwal1}).
This corresponds to superluminal light propagation.
\begin{figure}[htbp]
\centering{\vspace{0.5cm}}
\includegraphics[width=3.25in]{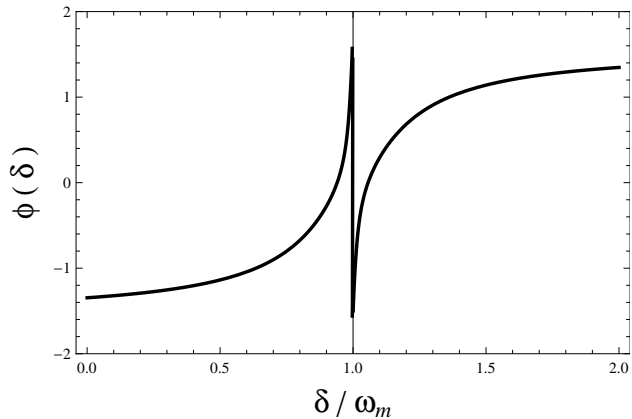}
\caption{Phase as a function of normalized frequency $\delta /
\omega_m$ for input coupling laser power $P_c = 1$ mW. The
parameters are the same as Fig. (\ref{fig2}).} \label{fig3}
\end{figure}
After some simplifications, we can write the ($2 \kappa
\tilde{c}_+$) in an instructive form. By using ($2 \kappa
\tilde{c}_+$) in Eq.(\ref{phase2}) and  we plot the phase which is
shown in Fig. (\ref{fig3}) as a function of the normalized
frequency $\delta / \omega_m$ for input coupling laser power $P_c
= 1$ mW. In these experimental parameters under no coupling field
the delay time is $\tau_0=1.48 \mu$sec. If there is no coupling,
the coupling constant is $\chi_0=0$. In the existence of coupling
EIT reverses the behavior of the system and the group delay
becomes negative. We plot the group delay $\tau$ as a function of
the pump power in Fig.(\ref{fig4}) which shows the group delay
$\tau$ as a function of the pump power $P_c$.

At low power pump group delay increases linearly with increasing
strength of the power pump for small power pump strengths. But
there is no transparency window if we use a pump power on the
order of $1-10 \mu$W. On the other hand, the group delay increases
slowly at large power of the coupling field. The probe pulse delay
can be tuned by calibrating the pump power in the probe resonance
($\delta=\pm \omega_m$) and $\Delta=\omega_m$. The pump power that
we have used in the Fig.(\ref{fig4}) is on the order of $10-400
\mu$W. The transparency window begin to be seen clearly.

\begin{figure}[htbp]
\centering{\vspace{0.5cm}}
\includegraphics[width=3.25in]{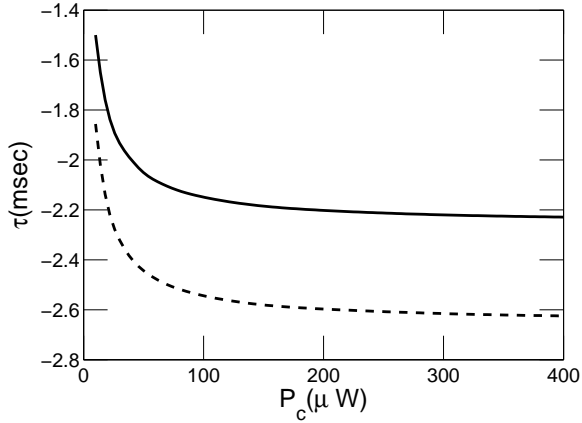}
\caption{Group delay as a function of the pump power. Time delay
of the probe in the presence of the coupling field as a function
of the power of the pumping or coupling field with
$\delta=\omega_m$. The solid and dashed lines are for
$\gamma_{m}=2\pi \times141$ Hz and $\gamma_{m}=2\pi \times120$ Hz,
respectively. The other parameters are the same with those of
Fig.\ref{fig2}.}. \label{fig4}
\end{figure}
We find large negative group delays of order $-2$ msec in a high
quality cavity under the action of a coupling laser and a probe
laser. In Fig.(\ref{fig4}) the group delay is negative, as a
result the fast light effect can be observed. This corresponds to
a superluminal situation. In Fig.(\ref{fig4}) the solid curve is
for $\gamma_{m}=2\pi 141$ Hz and the dashed curve is for
$\gamma_{m}=2\pi 120$ Hz. As the damped rate of the mirror
decreases, the group delay increases at all pump powers, however
after a critical value the group delays become approximately
constant.

%%%%%%%%%%%%%%%%%%%%%%%%%%%%%%%%%%%%%%%%%%%%%%%%%%%%%%%%%%%%%%%%%%%%%%%%%%%%%%%%%%%%%%%%%%%%%%%%%%%%%%%%%%%%%%%%%%%%%%%%%%%%%%%%%%%%
%%%%%%%%%%%%%%%%%%%%%%%%%%%%%%%%%%%%%%%%%%%%%%%%%%%%%%%%%%%%%%%%%%%%%%%%%%%%%%%%%%%%%%%%%%%%%%%%%%%%%%%%%%%%%%%%%%%%%%%%%%%%%%%%%%%%
\section{Conclusion}
\label{sec:concl}

We have examined the question of advance of the pulse under the
conditions of electromagnetically induced transparency in the
mechanical system contained in a high quality cavity. We have
investigated the tunable group delay of optical pulse in a
nanomechanical one-sided cavity system and also we have explored
the effect of the pump field power on the group delay. As the
strength of the pump power increases the group delay becomes
negatively larger.  However, at a critical value of pump field
power, group delay becomes approximately constant. Moreover, the
dependence of the phase of the transmitted probe pulse on the
frequency has been computed. In the absence of the coupling field,
the group delay time is $\tau_0=2/\kappa=1.48 \mu$sec. Whereas the
magnitude of the group delay is on the order of $-2$ msec at a
high pump power such as $400\mu$W for the parameters chosen in
Ref.\cite{aspel}. One can interpret that EIT reverses the behavior
of the nanomechanical system. Superluminal light can be observed
experimentally in a one-sided cavity system by adjusting the pump
power. Our results may be used for measuring the fast light in a
nanomechanical medium by comparing the group delay of the probe
pulse with a reference pulse propagating in the absence of medium.

%%%%%%%%%%%%%%%%%%%%%%%%%%%%%%%%%%%%%%%%%%%%%%%%%%%%%%%%%%%%%%%%%%%%%%%%%%%
%%%%%%%%%%%%%%%%%%%%%%%%%%%%%%%%%%%%%%%%%%%%%%%%%%%%%%%%%%%%%%%%%%%%%%%%%%%
\acknowledgments The author acknowledges support from Faculty
Support Program by the Council of Higher Education(Y\"{O}K) of
Turkey. The author thanks to Girish. S. Agarwal for many
stimulating, useful and fruitful discussion especially in this
work, to G. S. Agarwal for his hospitality during the visit for
three months at the Oklahama State University, Stillwater, USA, to
Kenan Qu and Sumei Huang for their fruitful discussions in Quantum
Optics.
%%%%%%%%%%%%%%%%%%%%%%%%%%%%%%%%%%%%%%%%%%%%%%%%%%%%%%%%%%%%%%%%%%%%%%%%%%%%%%%%%
%%%%%%%%%%%%%%%%%%%%%%%%%%%%%%%%%%%%%%%%%%%%%%%%%%%%%%%%%%%%%%%%%%%%%%%%%%%%%%%%%
%%%%%%%%%%%%%%%%%%%%%%%%%%%%%%%%%%%%%%%%%%%%%%%%%%%%%%%%%%%%%%%%%%%%%%%%%%%%%%%%%

\end{document}